\documentclass[aps,prb,twocolumn,a4paper,floatfix,showpacs]{revtex4-1}
\usepackage{amsmath,amssymb}
\usepackage{graphicx}
\usepackage{dcolumn}
\usepackage{bm}
\usepackage{subfigure}
\usepackage{soul}

\begin{document}

\title{Interfacial exchange interactions and magnetism of Ni$_2$MnAl/Fe bilayers}

\author{R. Yanes$^{1}$, E. Simon$^{2}$, S. Keller$^{1}$, B. Nagyfalusi $^3$, S. Khmelevsky$^{4}$, L. Szunyogh$^{2,5}$  and U. Nowak$^{1}$}
\affiliation{$^1$Department of Physics, University of Konstanz, Germany \\
             $^2$Department of Theoretical Physics, Budapest University of Technology and Economics, Budafoki \'{u}t 8., H-111 Budapest, Hungary \\
             $^3$ Institute for Solid State Physics and Optics, Wigner Research Centre for Physics, Hungarian Academy of Sciences, P.O. Box 49, H-1525 Budapest, Hungary\\
             $^4$Center for Computational Materials Science, Institute for Applied Physics, Vienna University of Technology, Wiedner Hauptstrasse 8, A-1060, Vienna, Austria \\
             $^5$MTA-BME Condensed Matter Research Group, Budafoki \'{u}t 8., H-111 Budapest, Hungary
}

\begin{abstract}
Based on a multi-scale calculations, combining ab-initio methods with spin dynamics simulations, we perform a detailed study of the magnetic behavior of  Ni$_2$MnAl/Fe  bilayers. Our simulations show that such a bilayer  exhibits a small exchange bias effect when the Ni$_{2}$MnAl Heusler alloy is in a disordered B2 phase. Additionally, we present an effective way to control the magnetic structure of the  Ni$_2$MnAl antiferromagnet, in the pseudo-ordered B2-I as well as the disordered B2 phases, via a spin-flop coupling to the Fe layer. 
\end{abstract}


\maketitle

\section{Introduction } 

Antiferromagnets build a class of materials which is used in magnetic multilayer devices, such as GMR sensors or magnetic tunnel junctions, to stabilize and control the magnetization of a ferromagnetic compound. This fact has increased the demand of antiferromagnets and it has led to an increasing interest in novel antiferromagnetic materials, with Heusler alloys as promising candidates for that \cite{hirohataIEEE15} . 

Heusler alloys are ternary inter-metallic compounds with the general formula X$_2$YZ, in which X and Y are typically transition metals and Z is an main group element. This kind of alloys has been in the center of intensive studies in the last decades, mainly due to the wide range of their multifunctional properties. These include magnetic shape memory effects, magneto-caloric and spintronic effects, as well as thermoelectric properties   amongst others. \cite{tanjaPSSC11}

Heusler alloys can be categorized in two distinct groups by their crystalline structures: half Heusler alloys with the form of XYZ in the C$_{1b}$ structure and full Heusler alloys with the form of X$_{2}$YZ in the L2$_{1}$ structure \cite{websterBook88}. The unit cell of the L2$_{1}$ structure consists of four interpenetrating face-centered cubic (fcc) lattices, while that of the C$_{1b}$ structure is formed by removing one of the X sites. The L2$_{1}$ structure  transforms into the so-called disordered B2 phase when the Y and Z atoms are mixed, replacing each other at random.  

The majority of magnetic Heusler alloys are ferromagnetic though it has been reported that some of them are ferrimagnets or even antiferromagnets. In particular, those compounds with 3d elements where only  the Mn atoms carry magnetic moments at Y site are antiferromagnets in the disordered B2 phase.
In this context Ni$_{2}$MnAl is especially interesting since it has been reported to exhibit an antiferromagnetic  behavior in the disordered B2 phase \cite{simonPRB15} as well as in the pseudo-order phase B2-I \cite{galanakisAPL11}. The latter is certain limit of the disordered B2 phase, where all the Mn atoms are located in the same (001) plane. Ni$_{2}$MnAl in the disordered B2 phase has a perfectly compensated antiferromagnetic ground state, where the Al and Ni atoms posses no net magnetic moment and the site and anti-site Mn atoms are equivalent \cite{simonPRB15}. 

Ni$_{2}$MnAl has been in the center of former studies regarding  shape memory applications \cite{acetJAP02,busgenPRB04,srivastavaAPL09} because of its capability to change its magnetic order along with its chemical order. The existence of exchange bias (EB) was reported for Heusler alloys which undergo  martensitic phase transitions  \cite{behlerAIPA13}, for Ru$_2$MnGe/Fe bilayers \cite{ballufJAP16} and, recently, Tsuchiya et al.\ published an experimental study of EB in Ni$_{2}$MnAl/Fe bilayers \cite{tsuchiyaJPDAP16}. In general, EB is related to the coupling between a ferromagnet (FM) and an antiferromagnet (AF) and its strength depends on the exchange interaction across the interface and the stability provided by the AF. This calls for a detailed study of the interfacial exchange interactions in Ni$_{2}$MnAl/Fe bilayers.

This paper is organized as follows: first we introduce a spin model which is based on first-principles calculations. Then, we analyze the exchange interactions  in the bulk and across the interface for the Ni$_2$MnAl(B2-I; B2)/Fe interfaces. In the next section, we present spin-dynamics simulations and analyze the possibility to control the magnetic state of the Ni$_2$MnAl layer via the Fe layer. We finish with a discussion of the origin of a small in-plane EB found in the Ni$_2$MnAl(B2)/Fe system.

\section{Model and numerical approach} 
In the following we study the magnetic properties of Ni$_2$MnAl(B2-I)/Fe and  Ni$_2$MnAl(B2)/Fe interfaces in the spirit of a multi-scale model, linking \textit{ab initio} calculations with dynamical spin model simulations. In terms of the fully relativistic Screened Korringa-Kohn-Rostoker (SKKR) Green's function method \cite{laszloPRB94, zellerPRB95,szunyoghPRB95} we perform self-consistent calculations of the Ni$_2$MnAl/Fe bilayers in the disordered local magnetic moment approach \cite{dlm-gyorffy}. We used the general gradient approximation (GGA) \cite{gga-2} in connection with the atomic sphere approximation and an angular momentum cut-off of $l_{max}=3$. We derive the exchange interactions between the magnetic moments by using the
spin-cluster expansion (SCE) technique \cite{sce-prb69, sce-prb72} that provides a systematic parametrization of the adiabatic energy of an itinerant magnetic system. Combining this method with the relativistic disordered local moment (RDLM) scheme \cite{rdlm-staunton-prl,rdlm-staunton-prb}, the parameters of the spin-Hamiltonian below can be determined on a quite general level \cite{laszloPRB11}. It is important to note that, due to the relativistic spin-orbit coupling, the exchange interactions between two spins form a $3 \times 3$ matrix. Furthermore, since the RDLM-SCE scheme relies on the paramagnetic state as reference, a priori knowledge of the magnetic ground state is not required, which makes it suitable for interface calculations.

 The magnetic properties of our system are well described by the following generalized spin model,
\begin{equation}\label{eq:Ham}
\mathcal{H}=-\frac{1}{2}\sum_{i,j}\vec{s}_{i}\boldsymbol{J}_{ij}\vec{s}_{j}
-\sum_{i}\vec{s}_{i}\boldsymbol{K}_{i}\vec{s}_{i}-\sum_{i}\mu_{i}\vec{H}_{A}\vec{s}_{i} \; ,
\end{equation}
where the $\vec{s}_{i}$ represent classical spins, i.~e. unit vectors along the direction of each magnetic moment at sites $i$. 
The first term stands for the exchange contribution to the energy, with $\boldsymbol{J}_{ij}$ denoting the tensorial exchange interaction between moment $i$ and $j$. The second term comprises the on-site anisotropy as well as the magneto-static energy, where $\boldsymbol{K}_{i}$ is called the anisotropy matrix.  In the presence of an external magnetic field, $\vec{H}_{A}$, the last term adds a Zeeman contribution to the Hamiltonian, where $\mu_{i}$ is the magnetic moment of the atom $i$.

The exchange tenors $\boldsymbol{J}_{ij}$ can be further decomposed into three parts,  $\boldsymbol{J}_{ij}=J_{ij}^{iso}\boldsymbol{I}+\boldsymbol{J}_{ij}^{S}+\boldsymbol{J}_{ij}^{A}$ \cite{udvardiPRB03},
 with the isotropic exchange interaction $J_{ij}^{iso}=\frac{1}{3} \mathrm{Tr}\bigl[\boldsymbol{J}_{ij}\bigr]$,  the traceless symmetric (anisotropic) part $\boldsymbol{J}_{ij}^{S}=\frac{1}{2}(\boldsymbol{J}_{ij}+\boldsymbol{J}_{ij}^{T})-J_{ij}^{iso}\boldsymbol{I}$, and the antisymmetric part $\boldsymbol{J}_{ij}^{A}=\frac{1}{2}(\boldsymbol{J}_{ij}-\boldsymbol{J}_{ij}^{T})$. The latter one is clearly related to the Dzyaloshinskii-Moriya (DM) interaction, $\vec{s}_{i}\boldsymbol{J_{ij}^{A}}\vec{s}_{j}=\vec{D}_{ij}\cdot(\vec{s}_{i}\times\vec{s}_{j})$,
with the DM vector $\vec{D}_{ij}$.  The DM interaction arises due to spin-orbit coupling and favors a perpendicular alignment of the spins $\vec{s}_{i}$ and $\vec{s}_{j}$ \cite{dzyaloshinskiiJPCS58, moriyaPR60}.

Our first principle calculations show that the Nickel as well as the Alumina atoms have negligible magnetic moments in both phases of the Ni$_{2}$MnAl compound, the pseudo-ordered B2-I phase as well as the disordered B2 phase. Therefore we restrict our spin dynamics analysis to the evolution of Fe and Mn moments only.

To study ground state properties along with spin dynamics at zero and finite temperatures we solve the stochastic Landau-Lifshitz-Gilbert (SLLG) equation,
\begin{equation}
	\frac{\partial \vec{s}_i}{\partial t}= -\frac{\gamma}{(1+\alpha^2)\mu_{\rm s}}\,\vec{s}_i \times \vec{H}_i -\frac{\gamma\alpha}{(1+\alpha^2)\mu_{\rm s}} \vec{s}_i \times\left(\vec{s}_i \times \vec{H}_i\right),
\end{equation}
by means of Langevin dynamics, using a Heun algorithm\cite{palaciosPRB98,nowakBook07}. The SLLG equation includes the gyromagnetic ratio $\gamma$, a phenomenological damping parameter, $\alpha$, and the effective field
\begin{equation}
	\vec{H}_i = {\vec{\zeta}_i(t)} - \frac{\partial\mathcal{H}}{\partial \vec{s}_i},
\end{equation}
which considers also the influence of a temperature $T$ by adding a stochastic noise term $\vec{\zeta}_i(t)$, obeying the properties of white noise\cite{lyberatosJPCM93} with
\begin{equation}
\langle{\vec{\zeta}_i(t)}\rangle=0,
\end{equation}
\begin{equation}
\langle\zeta_{i}^{\eta}(t)\zeta_{j}^{\theta}(t')\rangle=\frac{2k_{B}T\alpha\mu_{s}}{\gamma}\delta_{ij}\delta_{\eta\theta}\delta(t-t').
\end{equation}
Here $i,\;j$ denote lattice sites and $\eta$ and $\theta$ Cartesian components of the stochastic noise.

\section{Results and discussions} 

\subsection{Ab initio results} 

\begin{figure}[h]
 	(a)\includegraphics[width=0.4\textwidth]{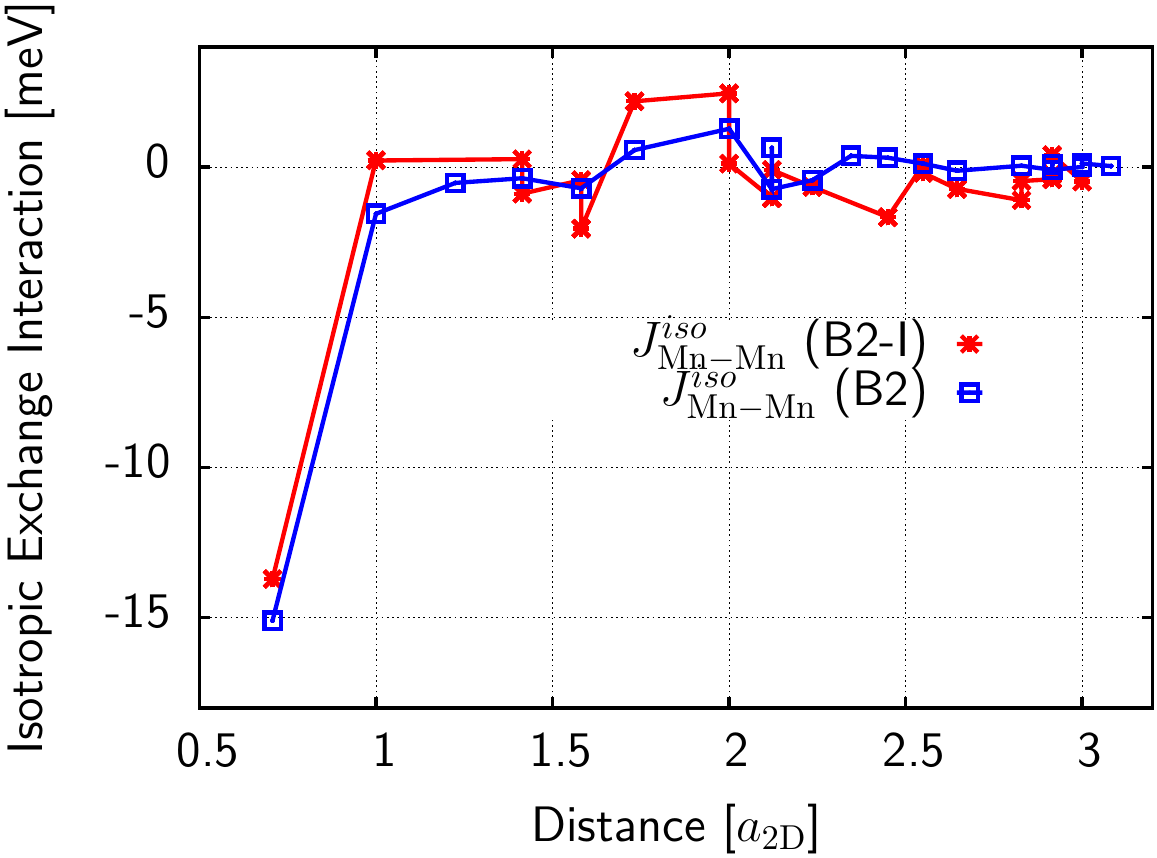}
     (b)\includegraphics[width=0.4\textwidth]{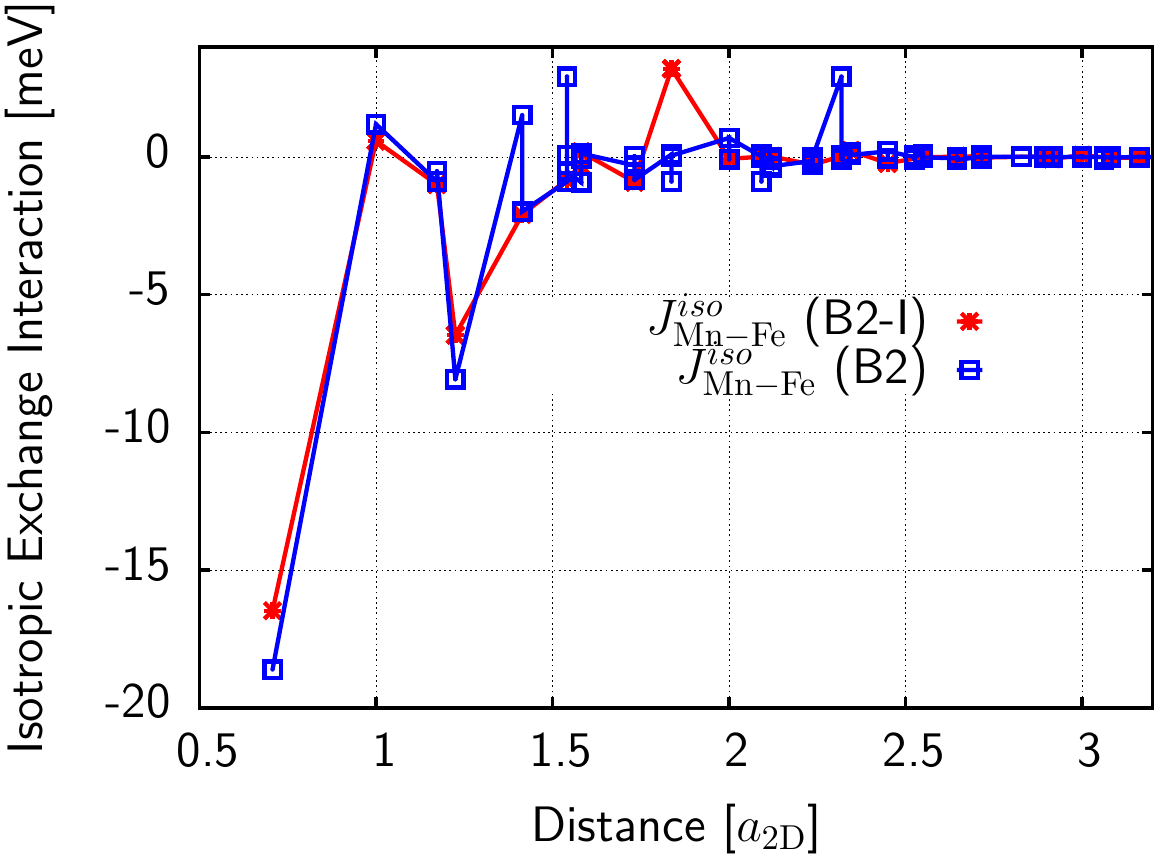}
 	\caption{(Color online). Isotropic exchange interaction as a function of pair distance, (a) between Mn-Mn atoms in B2 and B2-I phases of Ni$_{2}$MnAl bulk,  (b) across the interface between Mn-Fe atoms in Ni$_{2}$MnAl(B2-I)/Fe  and  Ni$_{2}$MnAl(B2)/Fe bilayers.
 	}
 	\label{fig:Jiso}
 \end{figure}

For the two cases investigated in this work, the Ni$_{2}$MnAl(B2-I)/Fe  and  Ni$_{2}$MnAl(B2)/Fe bilayer, we first calculated the exchange interactions, the magnetic moment and the on-site anisotropy layered resolved  with the methods described above.  In Fig.$\,$\ref{fig:Jiso} the isotropic contribution of the exchange interaction between Mn-Mn and Mn-Fe neighbors are presented as a function of the distance between spin pairs. 
For the isotropic Mn-Mn exchange interactions our results indicate a similar behavior for the pseudo-ordered  B2-I and the disordered B2 phase. The dominant nearest neighbor Mn-Mn exchange interaction, $J_{1,{\rm Mn-Mn}}\approx-15\,$meV, supports antiferromagnetic order while the magnitude of the exchange interactions between Mn atoms in successive shells decay rapidly (Fig. \ref{fig:Jiso}(a)). 

The exchange interactions between Mn and Fe atoms across the interface are plotted in Fig. \ref{fig:Jiso}(b). 
The dominant Mn-Fe exchange interaction is again the nearest neighbor one, favoring antiferromagnetic alignment. Remarkably, this interaction is even larger in magnitude than the nearest neighbor Mn-Mn interaction in the bulk. It should also be mentioned that the magnitude of the nearest neighbor exchange interaction in bulk Fe,  $J_{1,{\rm Fe-Fe}}\approx 50\,$meV, is again much larger in magnitude than the above interactions. A summary of the most relevant isotropic exchange parameters is given in Table \ref{tab:mag_par}.

Another important parameter, which influences the magnetic behavior of a magnetic bilayer and which can lead to the existence of exchange bias is the magnetic anisotropy energy (MAE). It has been reported \cite{simonPRB15} that bulk Ni$_{2}$MnAl(B2-I) has a small in-plane anisotropy with a magnitude of  $0.19\,$meV per spin, while in the case of the perfectly disordered B2 phase the MAE is negligible. Close to the AF/FM  interface, however, the magnetic anisotropy is modified. In case of the Ni$_{2}$MnAl(B2-I)/Fe interface the preferred magnetic orientation is in-plane with an energy of $0.03$ meV in the interface Mn layer and $0.06$ meV in the Fe layer. Similarly, an easy plane anisotropy was determined for the Ni$_{2}$MnAl(B2)/Fe interface, with a MAE of  $0.05$ meV and $0.10$ meV in the interface Mn and Fe layers, respectively. 

\begin{table}
	\caption{\label{tab:mag_par}Calculated maximum isotropic exchange interactions between nearest neighbors, $J_{ij}^{\rm {iso}}$ 
	 (in meV) and magnitudes of the magnetic moment.}
	\begin{ruledtabular}
		\begin{tabular}{lccccc}
		Material& $J^{iso}_{1,\rm{Mn-Mn}}$& $J_{1,\rm(Mn-Fe)}^{iso}$&  $J^{iso}_{1,\,\rm{Fe-Fe}}$& $\mu_{Mn}$& $\mu_{Fe}$\\
			\colrule
{\footnotesize Ni$_{2}$MnAl(B2-I)/Fe}	&-13.71&-16.45&52.62&3.32&2.4\\
{\footnotesize 	Ni$_{2}$MnAl(B2)/Fe}	&-15.1&	-18.61&52.67&3.35&2.45\\
		\end{tabular}
	\end{ruledtabular}
\end{table}%


\subsection{Spin dynamics simulations} 
For our spin-dynamics simulations we use the model parameters as determined above from first principles. We suppose the Ni and Al atoms to be nonmagnetic and only consider the dynamics of the Mn and Fe moments.
 The antiferromagnet  is hence modeled by the Mn sub-lattice, forming in total $30\times30\times t_{\rm{AF}}$ unit cells and the ferromagnet by $30\times30\times 3$ unit cells, $t_{\rm{AF}}$ denoting the number of Ni$_2$MnAl atomic monolayers perpendicular to the interface (in the following labeled [Ni$_2$MnAl(B2I;B2)]$_{t_{\rm{AF}}}$/[Fe]$_3$). We consider open boundary conditions.
 
For the case of the disordered B2 phase, the Mn atoms are statistically distributed. The magnitudes of the magnetic moments of Mn and Fe atoms were taken uniformly in the sample using the values given in Table \ref{tab:mag_par}. Additionally we approximate the effects of the magneto-static interaction in the FM layer as an uniaxial shape anisotropy with $K_{\rm{Fe}}=-0.134$ meV and the magnetic hard axis perpendicular  to the FM/AFM interface. 
 
 In the following sections we will analyze the magnetic properties of the two types of bilayers described above. We evaluate the in-plane hysteresis loops and explore the existence of EB and the switching of the magnetic structure of the Ni$_2$MnAl layer. 

\subsubsection{Hysteresis in the pseudo-ordered Ni$_{2}$MnAl(B2-I)/Fe bilayer} 
To study the magnetic behavior of this bilayer we calculate hysteresis loops as a succession of quasi-equilibrium states determined by the numerical integration of the SLLG equation applied to the spin model described above. The tensorial exchange interactions are considered up to 11th neighbor. Initially we prepare the system similar to experiments by simulating a field-cooling process. This process starts from a random spin configuration in the AFM at an initial temperature $T$ above the N\'{e}el temperature of the AFM but below the Curie temperature of the FM, and proceeds to a final temperature  under the influence  of  an in-plane magnetic $H_{\rm{FC}}$.

\begin{figure}[h]
 	\subfigure{\label{fig:B21Fe-Conf}(a)\includegraphics[width=0.35\textwidth]{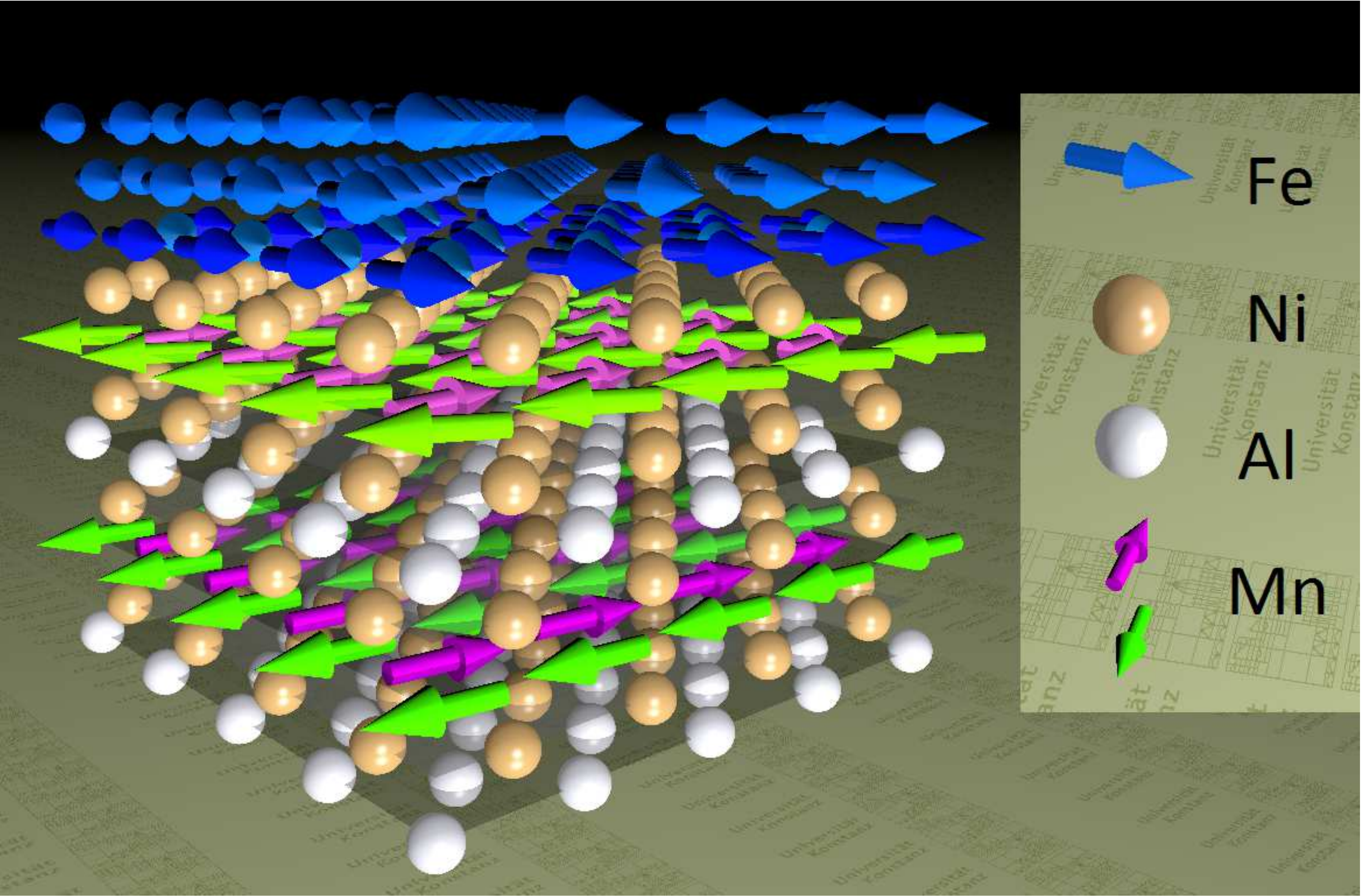}}
 	\subfigure{\label{fig:B21Lz1n}(b)\includegraphics[width=0.205\textwidth]{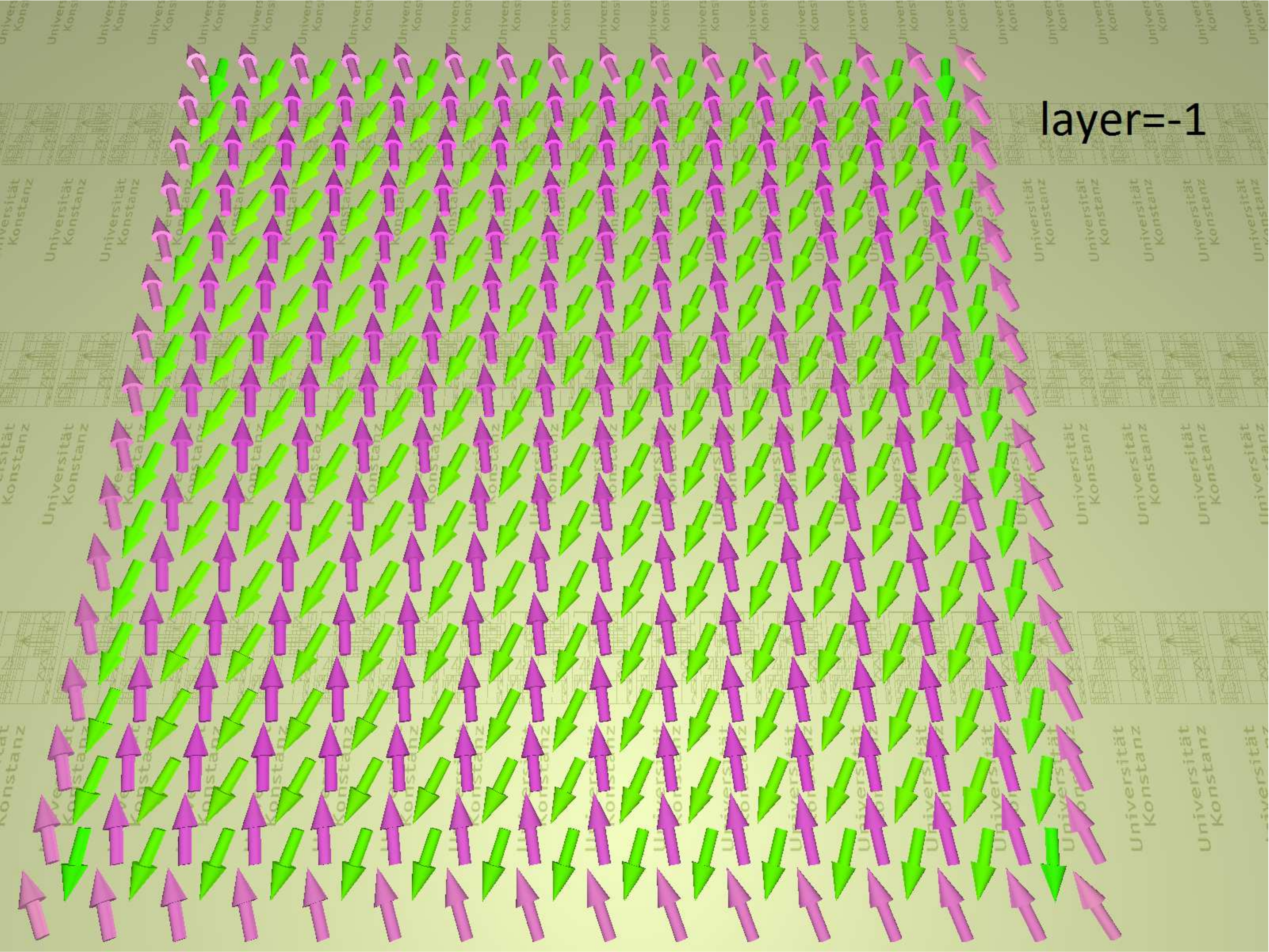}}
 	 \subfigure{\label{fig:B21Lz5n}\includegraphics[width=0.202\textwidth]{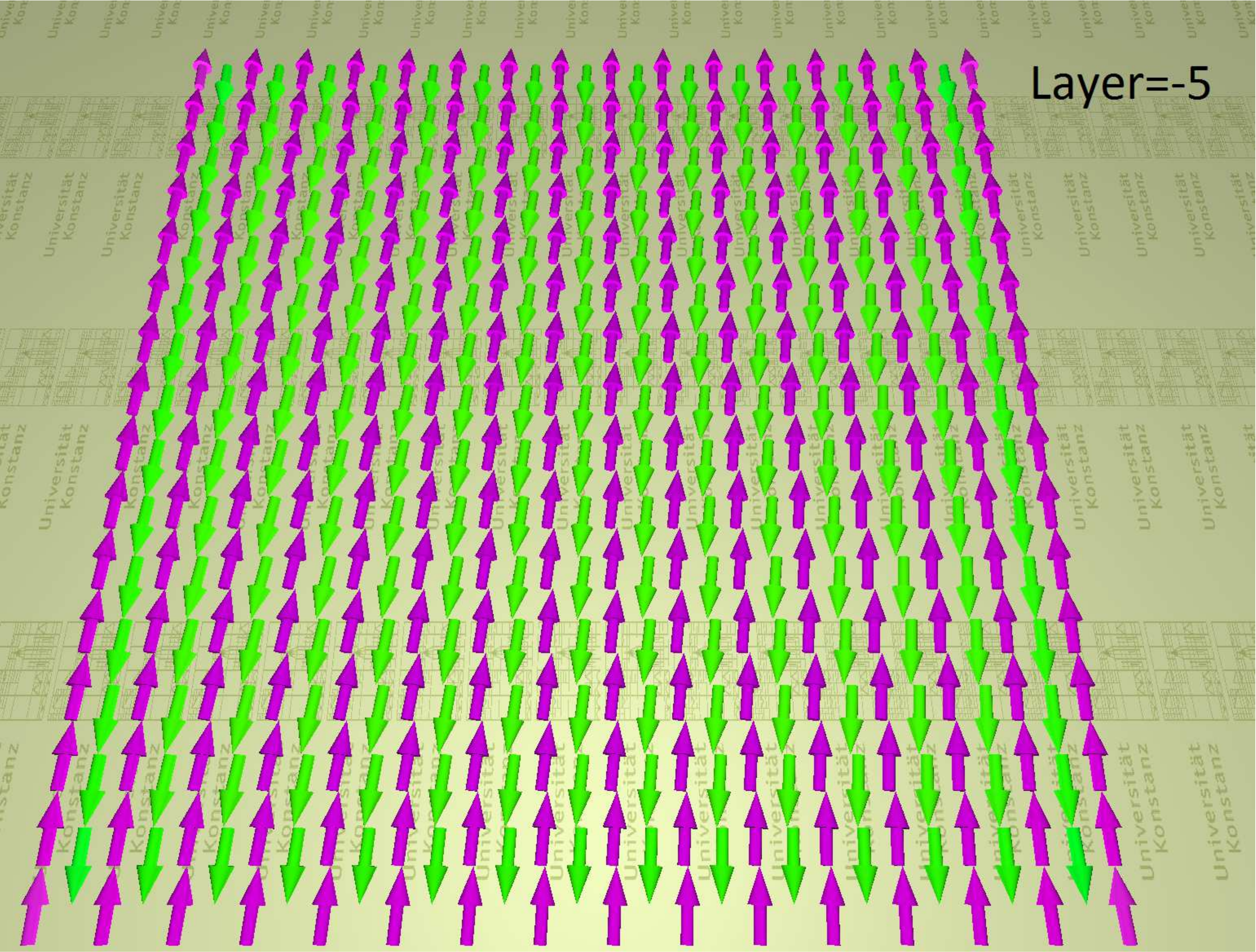}}
   	\caption{(Color online).(a) Sketch of the magnetic state of the Ni$_{2}$MnAl(B2-I)/Fe interface after the field cooling process.  (b) Spin configurations of the first two Mn layers starting at the interface to the Fe (interface Ni layer has layer index 0). 
	} 
	\label{fig:MagSB21Fe}
 \end{figure}

After the field cooling process Mn as well as Fe magnetic moments are oriented in-plane which is in correspondence with the calculated in-plane magnetic anisotropy. Importantly, the direction of the Mn moments is nearly perpendicular to that of the Fe moments, a consequence of the so-called spin-flop coupling \cite{KoonPRL97}. 
Near the interface, the Mn moments are slightly tilted from this perpendicular ($x$-) direction, leading to a very small net magnetic moment, anti-parallel to the Fe moments. This configuration follows from the strong antiferromagnetic  exchange interaction between Mn-Fe moments and the fact that the  interface between Ni$_{2}$MnAl(B2-I)/Fe is compensated (equal number of Mn moments in both magnetic sub-lattice). This spin configuration is shown in Fig. \ref{fig:MagSB21Fe}.

\begin{figure}[h]
  	(a)\includegraphics[width=0.4\textwidth]{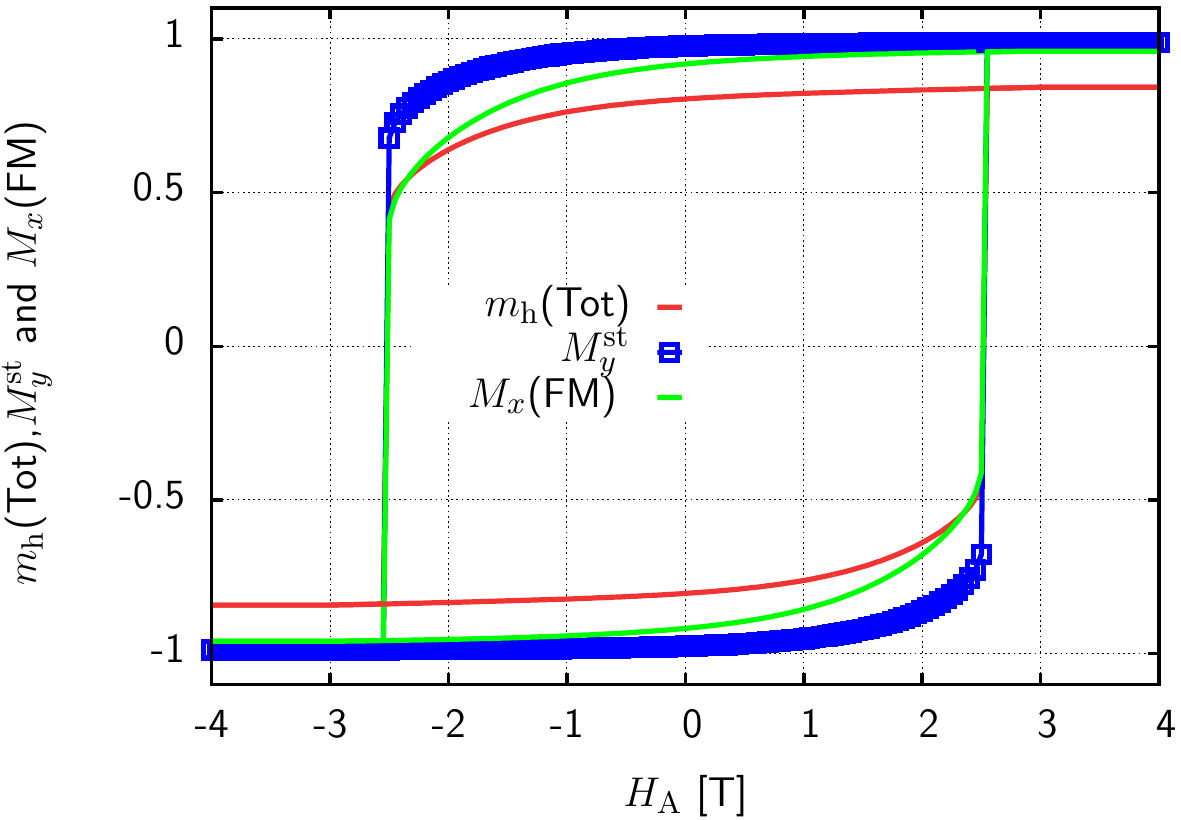}
     (b)\includegraphics[width=0.4\textwidth]{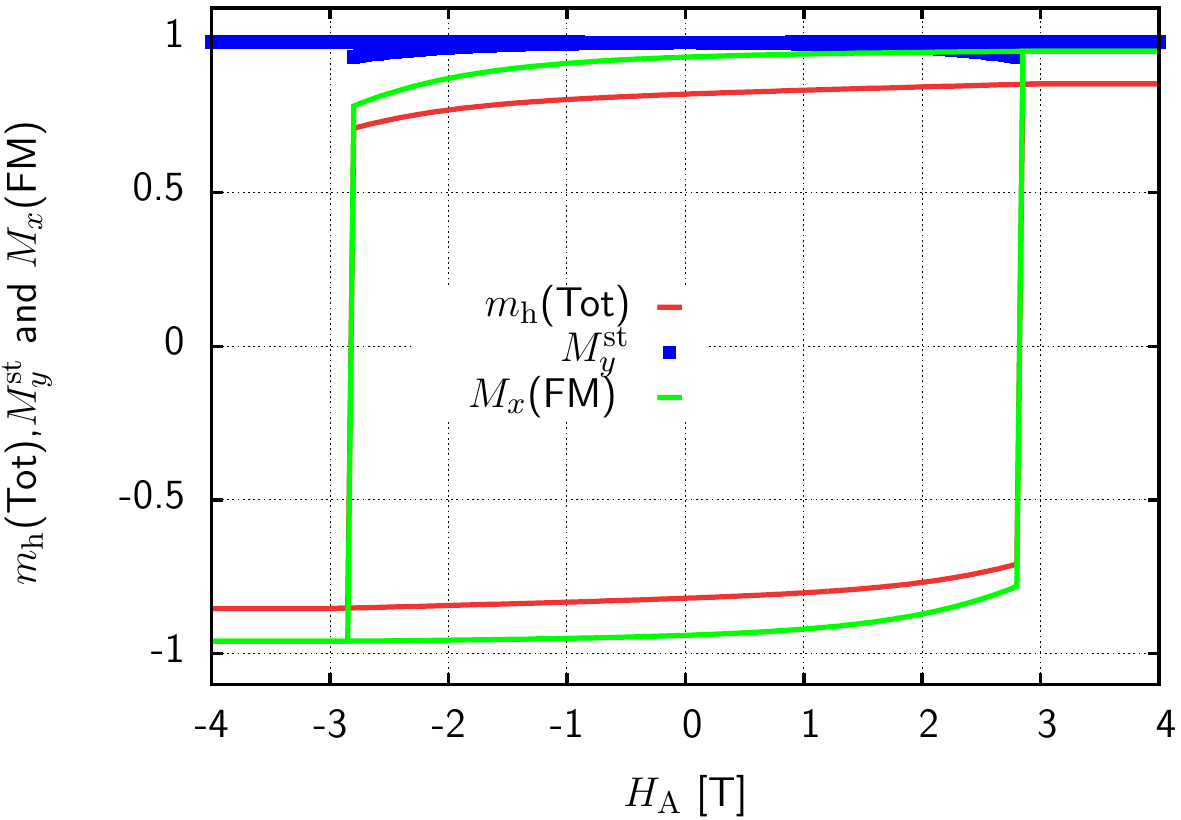}
 	\caption{(Color online). In-plane hysteresis loops for a [Ni$_2$MnAl(B2-I)]$_{t_{\rm{AF}}}$/[Fe]$_3$ bilayer. (a) thickness of the AF  $t_{\rm{AF}}=20a_z$ and (b) $t_{\rm{AF}}=30a_z$. Shown are the normalized magnetization of the FM along the applied field direction, $M_x$(FM), and the total magnetization of the system along the direction of the applied field, $m_{\rm{h}}$(Tot), as well as the in-plane normalized antiferromagnetic order parameter $M^{\rm{st}}_y$ perpendicular to the applied field. 
 	}
 	\label{fig:OPB21Fe}
 \end{figure}

We investigate the  switching mechanism  and the possible existence of EB for different values of the thickness $t_{\rm{AF}}$ of the Ni$_{2}$MnAl(B2-I) layer. Our findings indicate that for perfect bilayers there is no EB within our numerical error of $\pm 75$Oe. However, the behavior of the AF changes drastically as $t_{\rm{AF}}$ is increased. As an example in Fig. \ref{fig:OPB21Fe} we show hysteresis loops for two different thicknesses of the AF, focusing on the evolution of the magnetization of the FM along the direction of the applied field, $M_x$(FM),  the total magnetization of the system along the direction of the applied field, $m_{\rm{h}}$(Tot), as well as the in-plane antiferromagnetic order parameter, $M^{\rm{st}}_y$, perpendicular to the applied field. 

We observe that during hysteresis the Fe moments rotate coherently, staying mostly in-plane. 
For the smaller thickness, due to the strong exchange interactions between Mn-Fe moments, the small net magnetic moment of the AF close to the interface also rotates,  maintaining the antiferromagnetic order. Finally the AF switches following the FM (see Fig. \ref{fig:OPB21Fe} (a)). 
When the thickness of the AF increases, and concomitantly the relevance of the on-site MAE of the AF, the AF cannot switch anymore and the antiferromagnetic order parameter, $M^{\rm{st}}_y$, remains close to unity (see Fig. \ref{fig:OPB21Fe} (b)). Nevertheless, the small canting of the Mn moments at the interface switches with the FM so that the magnetic moment of the AF maintains its direction antiparallel to the Fe moments. 

These results indicate that for sufficiently thin layers it is possible to manipulate the magnetic order of the antiferromagnetic Ni$_2$MnAl layer through the magnetization of the Fe layer. A similar control of the AF magnetization by the FM layer has been reported for NiFe/IrMn/MgO/Pt heterostructure\cite{parkNatMat11} as a key point to the use of that system in an AF-based tunnel junction. Our finding opens hence the door for new Heusler-alloy-based antiferromagnetic spintronic devices.
  
\subsubsection{Hysteresis in the disordered Ni$_{2}$MnAl(B2)/Fe bilayer} 
\begin{figure}
 \subfigure{\label{fig:B2Fe-Conf}(a)\includegraphics[width=0.35\textwidth]{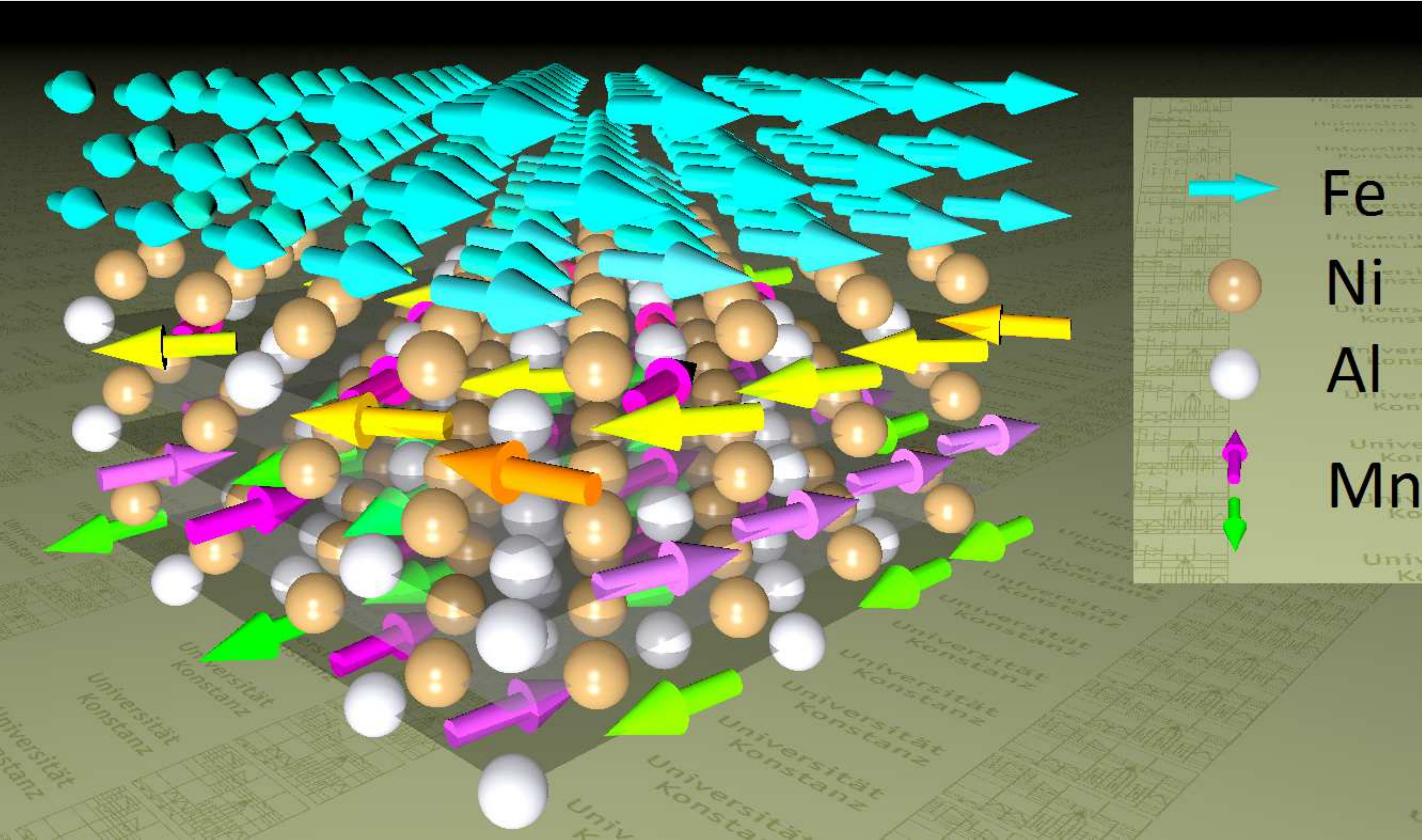}}
 \subfigure{\label{fig:Lz1n}(b)\includegraphics[width=0.21\textwidth]{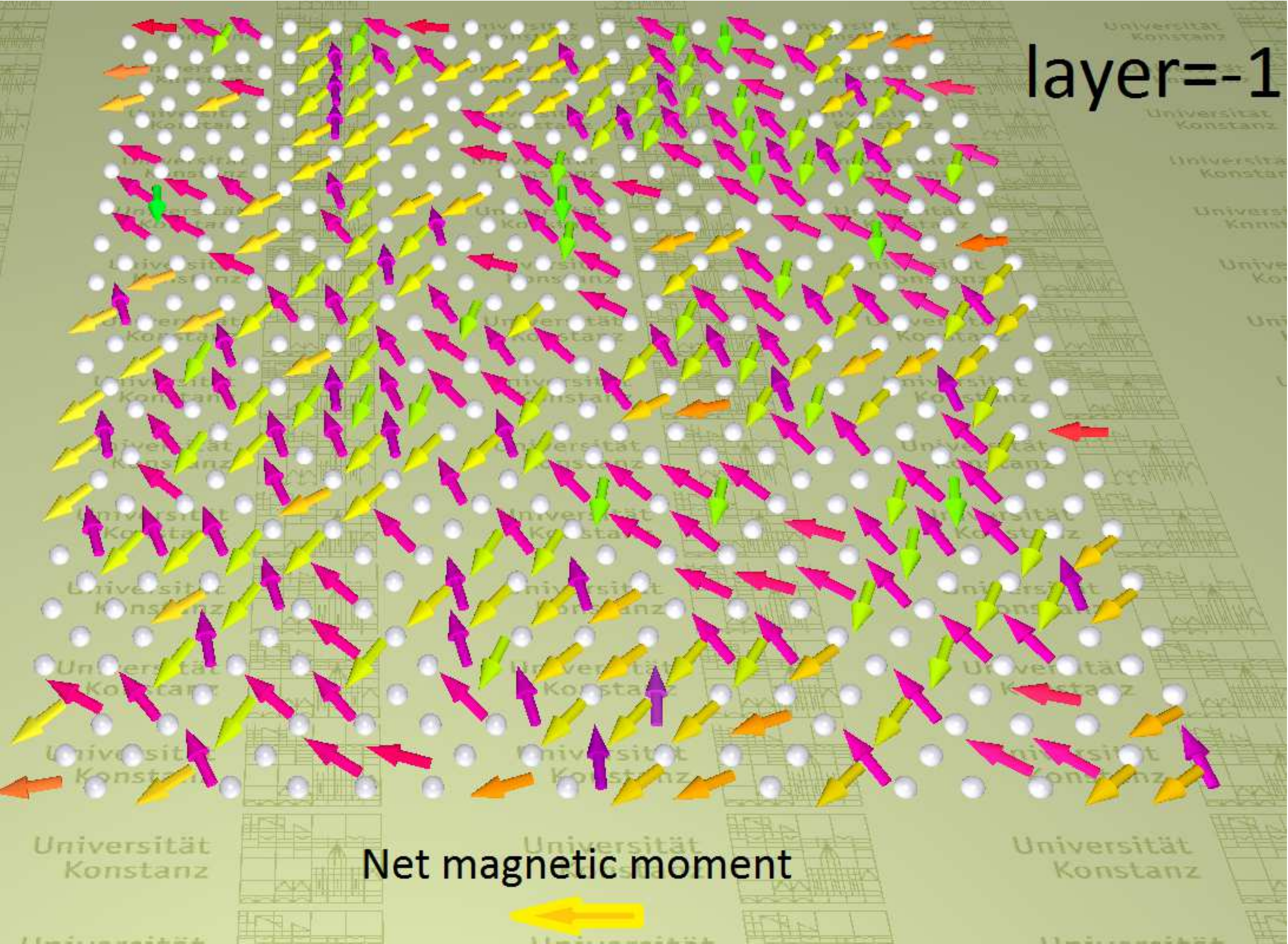}}
  \subfigure{\label{fig:Lz3n}\includegraphics[width=0.205\textwidth]{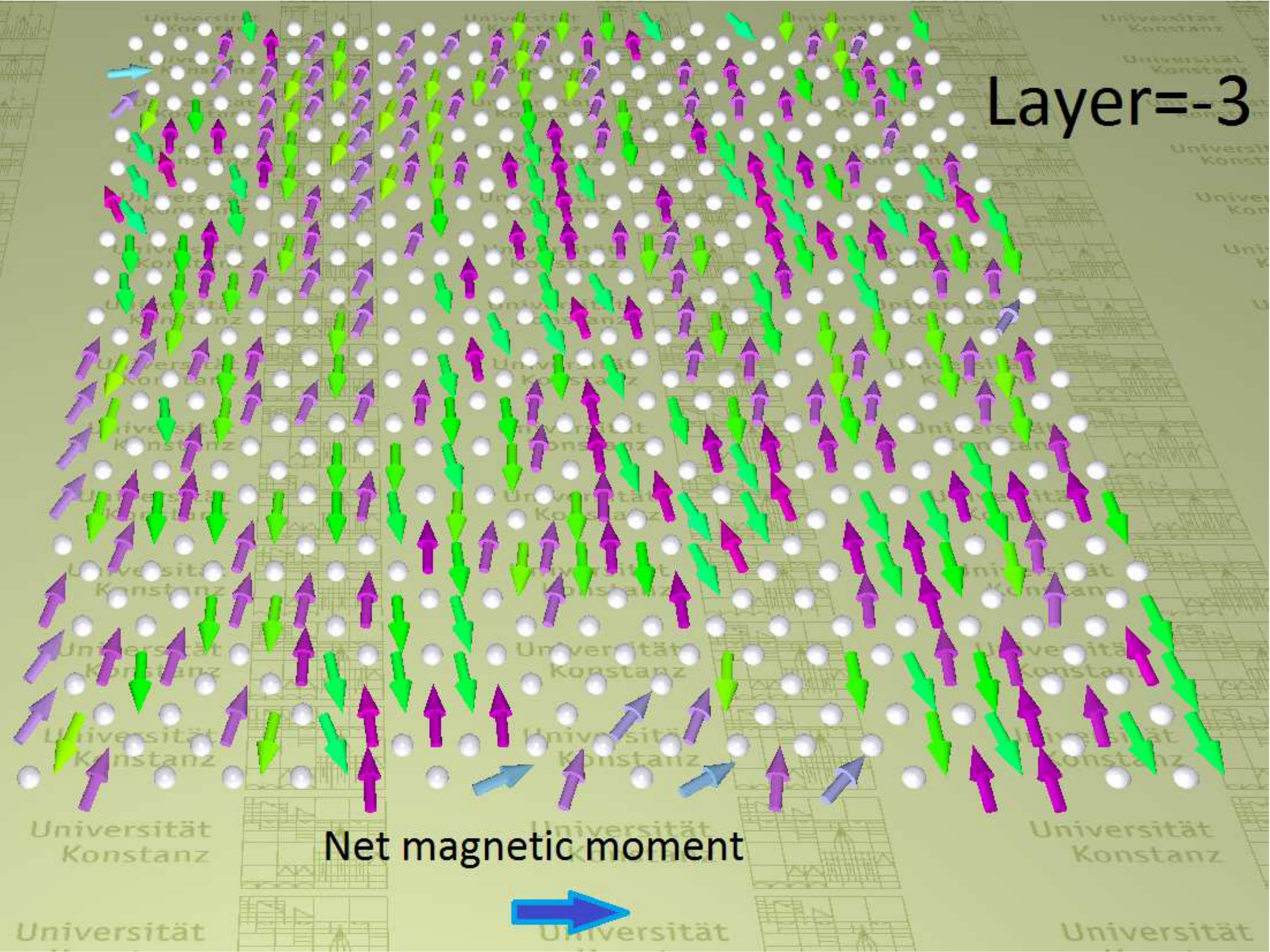}}
   
 \caption{(Color online).(a) Sketch of the magnetic state of the [Ni$_2$MnAl(B2)]$_{10}$/[Fe]$_3$ interface after the field cooling process. (b) Spin configurations of the first two Mn layers starting at the interface to the Fe (interface Ni layer has layer index 0). 
 } 
 \label{fig:B2Fe-FC}
\end{figure}

Calculations similar to the ones described above were performed for the disordered Ni$_2$MnAl(B2)/Fe system. First of all, it is important to note that, as a result of the chemical disorder in the B2 phase and its low effective anisotropy, much more complex spin structures appear in the AF after the field cooling process (see Fig.\ \ref{fig:B2Fe-FC}). As before, the Fe moments are aligned along the $x-$direction, the direction of the field during cooling. Again, we observe a kind of spin-flop coupling with the AF ordered mostly perpendicular to the FM and in-plane. However, the canting of the Mn moments at the interface is much more pronounced as compared to the  Ni$_2$MnAl(B2-I)/Fe system (see the red and yellow Mn moments in the first layer of Fig.\ \ref{fig:B2Fe-FC} b)). The reason for this much stronger canting is the structural disorder in Mn moment positions. Due to the statistical distribution of the Mn moments with some probability clusters of moments within the same sub-lattice appear. In these clusters the moments have a smaller connectivity to Mn moments of the other Mn sub-lattice where the antiferromagnetic exchange would counteract the canting. As a consequence, larger tilting angles and with that a larger net magnetization antiparallel to the Fe magnetization appears. However, the effective coupling between Mn and Fe layers is still smaller since only 50 \% of the sites of the layer which is closest to the Fe are occupied. For comparison, in the pseudo-ordered  Ni$_2$MnAl(B2-I)/Fe interface 100\% of these sites are occupied by Mn atoms.

\begin{figure}
 \subfigure{\label{fig:MagOrd}(a)\includegraphics[width=0.4\textwidth]{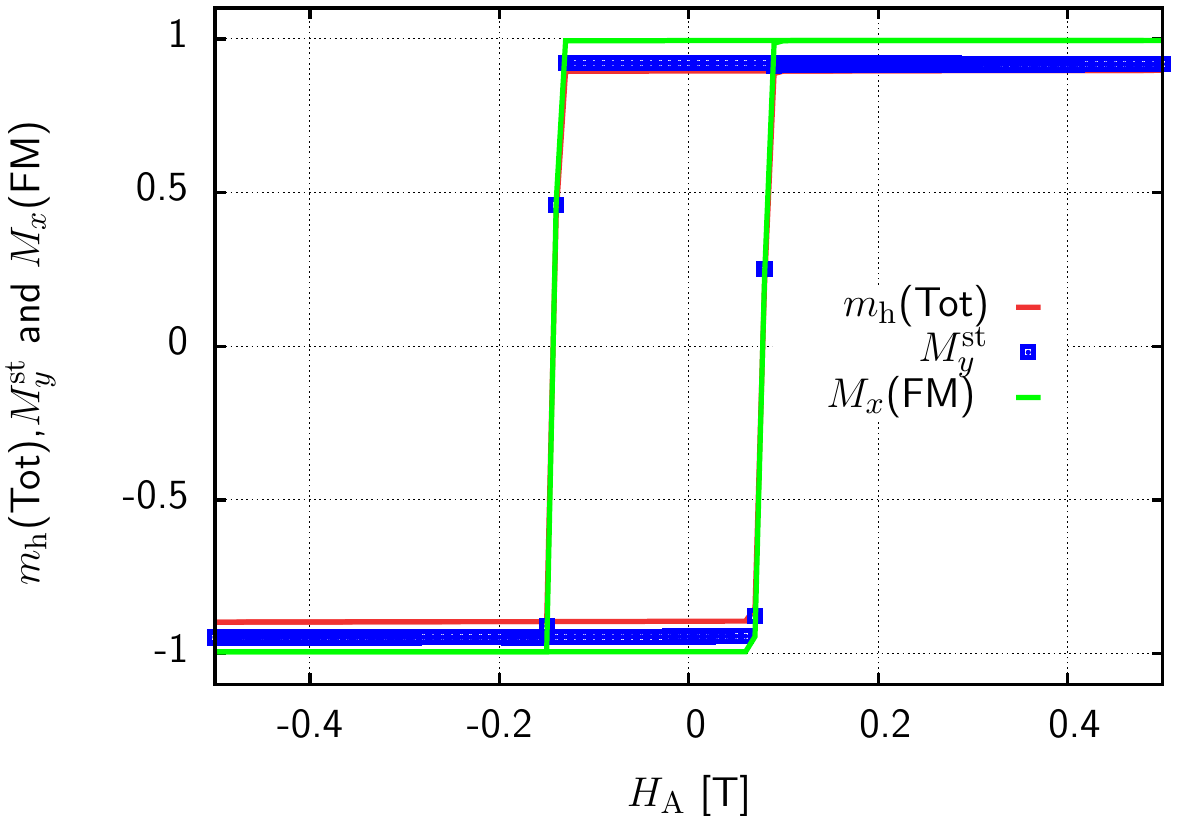}}
 \subfigure{\label{fig:hystC}(b)\includegraphics[width=0.4\textwidth]{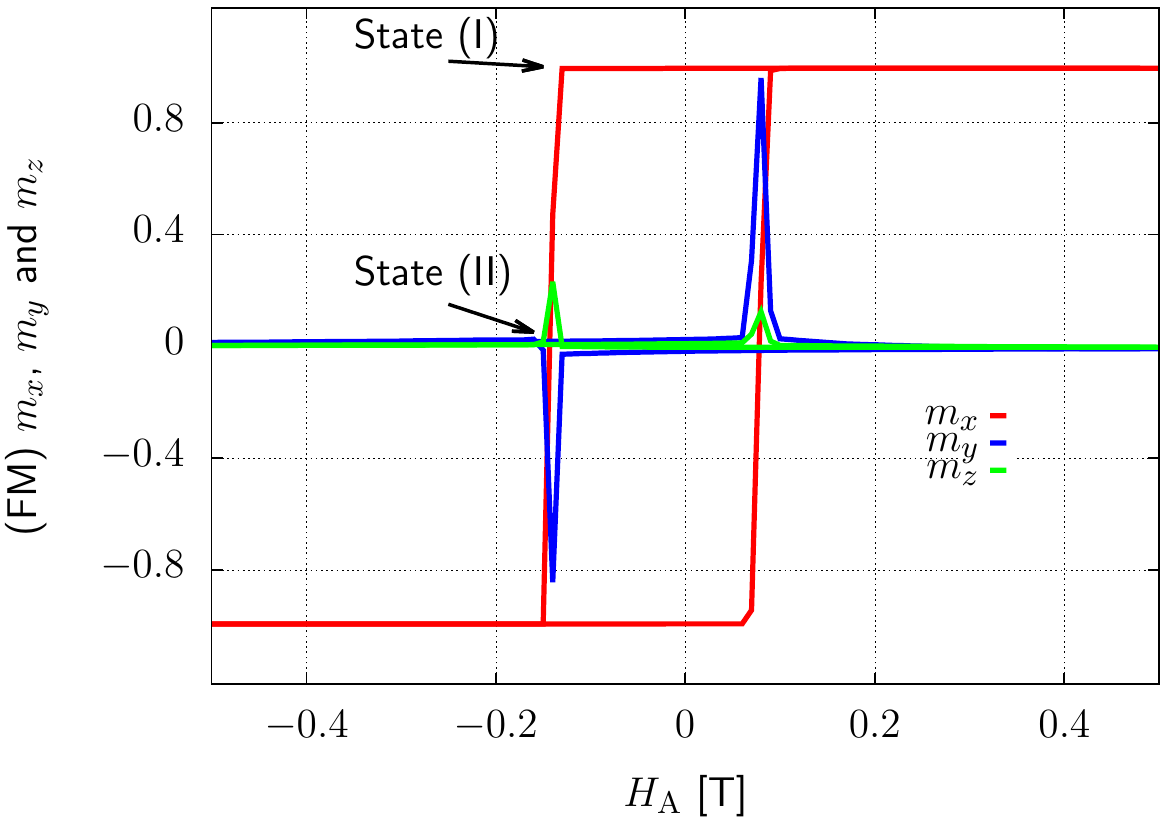}}
 \subfigure{\label{fig:conf1}(c)\includegraphics[width=0.215\textwidth]{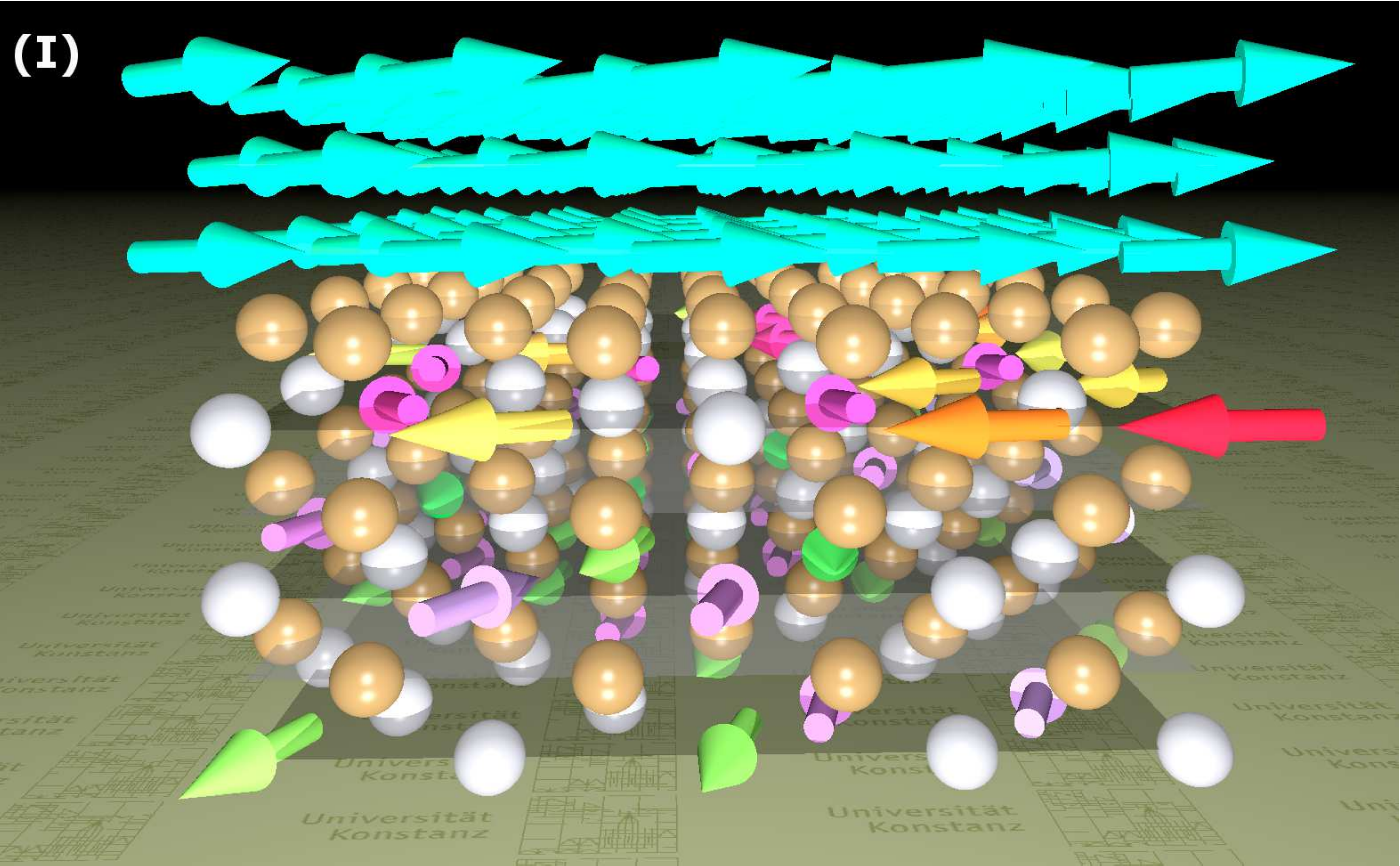}}
  \subfigure{\label{fig:conf59}\includegraphics[width=0.205\textwidth]{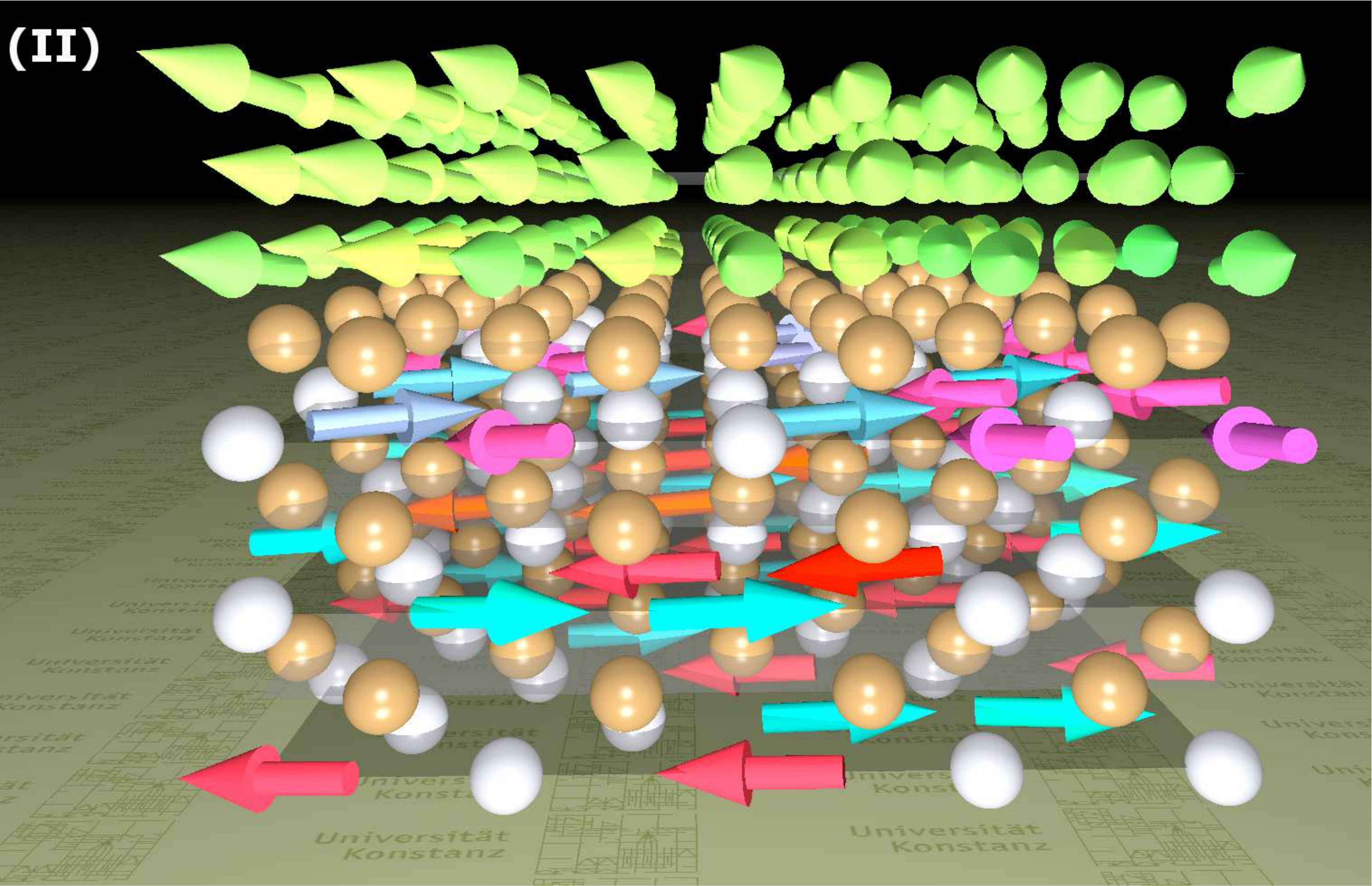}}
 \caption{(Color online). (a) In-plane hysteresis loops for a disordered [Ni$_2$MnAl(B2)]$_{10}$/[Fe]$_3$  bilayer. Shown are the normalized magnetization of the FM, $M$(FM), and the total magnetization of the system along the direction of the applied field, $m_{\rm{h}}$(Tot) as well as the in-plane normalized antiferromagnetic order parameter, $M_{\rm{st}}$(AF), perpendicular to the applied field. (b) Components of the Fe magnetization. (c) Sketches of the magnetic state of the Ni$_{2}$MnAl(B2)/Fe bilayer before (I)  and during  (II) the switching process.} \label{fig:HystxB2Fe}
\end{figure}

In Fig. \ref{fig:HystxB2Fe} (a) hysteresis curves are presented. These hysteresis loops are shifted horizontally, corresponding to an exchange bias field of $H_{\rm{EB}}=200 \pm 50$Oe. From the difference between the magnetization of the FM and the total magnetization one can see that not only the Fe moments contribute to the hysteresis loops but also the Mn moments. Furthermore, the sub-lattice magnetization of the AF switches as well indicating that the AF follows the FM as in the case of the pseudo-ordered bilayer for the thin AF layer (see Fig.\
\ref{fig:HystxB2Fe} (c)).

During the switching process the Fe moments rotate again mainly in plane, as we can see in Fig.\ \ref{fig:HystxB2Fe}(b) where the $x,\; y$ and $z$ components of the Fe magnetization are plotted. However a small out-of-plane component of the magnetization appears as well during the switching in both branches of the hysteresis loops. 
The coercive field is much smaller than in the previous case of the pseudo-ordered bilayer. This smaller coercive field is due to the fact that the effective interface coupling is smaller because of the smaller occupancy with magnetic Mn atoms at the interface. Furthermore, the anisotropy of the AF is smaller which leads to a smaller stability of the AF against switching.  

\begin{figure}
	\includegraphics[width=0.4\textwidth]{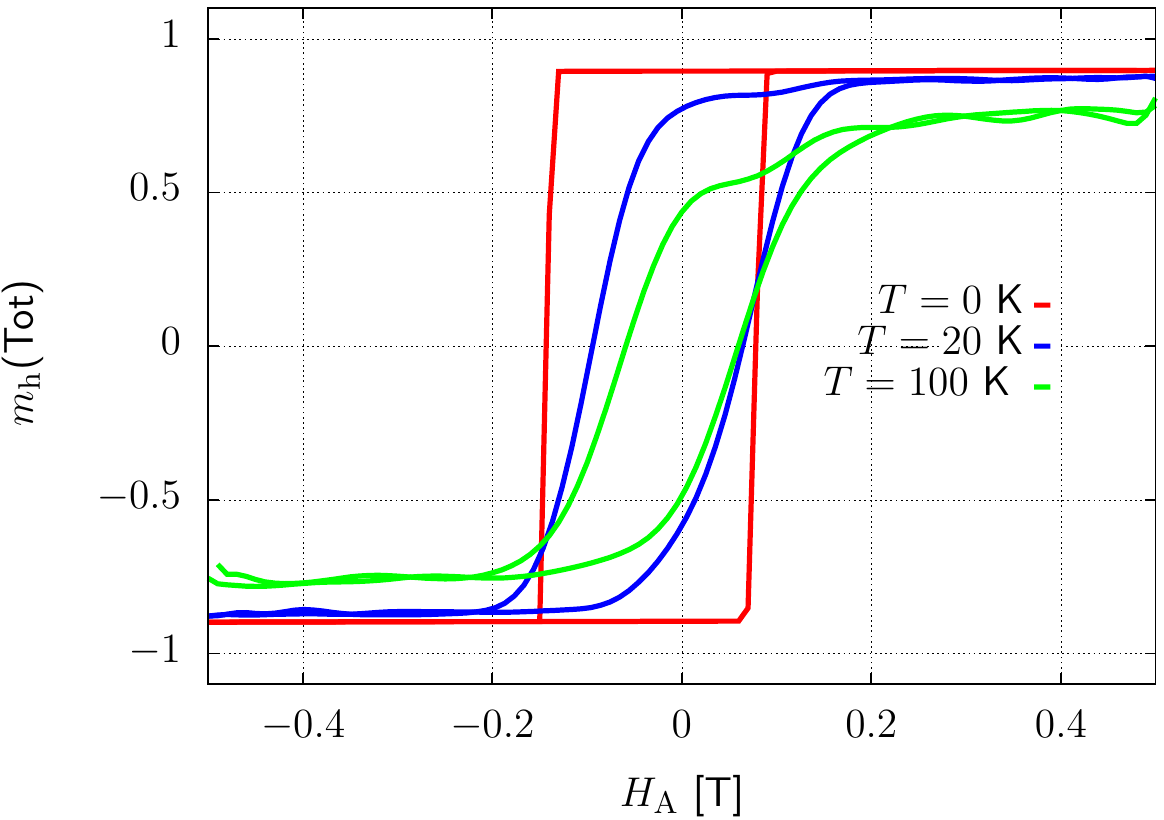}
	\caption{(Color online). In-plane hysteresis loops for a disordered [Ni$_2$MnAl(B2)]$_{10}$/[Fe]$_3$ bilayer at different temperatures.	}
	\label{fig:HebTemp}
\end{figure}

For an investigation of the thermal stability of the EB effect mean hysteresis loops where calculated as an  average over 5 hysteresis loops performed using the same spacial distribution of Mn-Al atoms. The EB we find is not only rather small and but also unstable against thermal fluctuations (see Fig.\ \ref{fig:HebTemp}).
Our results suggest a blocking temperature below 100K. 

Over all our simulations indicate that the EB is related to the disorder --- the lack of perfect compensation due to the random distributions of the Mn and Al atoms into the  Y-Z positions in the Heusler alloy --- in combination with the anisotropy in the AF. As a consequence a small part of the interface magnetization of the AF becomes frozen and does not switch with the FM which leads to the EB. This conclusion is supported by the fact that the EB vanishes for increasing lateral size.
 
\section{Conclusion} 
In summary, by means of a multi-scale modeling we investigate the interfacial magnetic interactions, the magnetic state and  the hysteresis loops of Ni$_2$MnAl(B2-I;B2)/Fe bilayers. Based on first principles calculations we find a strong negative Mn-Fe interface interaction, exceeding the antiferromagnetic interactions within the  Ni$_2$MnAl. For the disordered Ni$_2$MnAl(B2)/Fe bilayer we find a small EB at low temperatures in agreement with recent measurements \cite{tsuchiyaJPDAP16}. The existence of such an exchange bias is related to the disorder in the AF and with that to a lack of perfect compensation at the interface. More importantly, we have shown that it is possible to switch the magnetic structure of the antiferromagnetic Ni$_2$MnAl layer in both, the pseudo-ordered B2-I and disordered B2 phase, via a spin-flop coupling to the ferromagnetic Fe capping layer. This open the doors for the control of antiferromagnetic Heusler alloys in spintronic devices with antiferromagnetic components.

\begin{acknowledgements}
 This work was supported by the European Commission via the Collaborative Project HARFIR (Project No. 604398) and the National
Research, Development and Innovation Office of Hungary under Project No. K115575. 
\end{acknowledgements}

\end{document}